\documentclass[twocolumn,showpacs,aps,prl]{revtex4}
\usepackage{dcolumn}
\usepackage{epsf}
\begin{document}
\title{First-principles theoretical evaluation of crystalline zirconia
and hafnia  as gate oxides for  Si microelectronics} 
\author{Vincenzo  Fiorentini}
\affiliation{INFM and Dipartimento di  Fisica, Universit\`a di Cagliari,
Cittadella  Universitaria, I-09042 Monserrato (CA), Italy}
\author{Gianluca Gulleri}
\affiliation{INFM and Dipartimento di  Fisica, Universit\`a di Cagliari,
Cittadella  Universitaria, I-09042 Monserrato (CA), Italy}
\date{\today} 
\begin{abstract}
Parameters determining the  performance
 of the crystalline   oxides zirconia 
(ZrO$_2$) and hafnia (HfO$_2$)
 as gate insulators  in nanometric Si electronics are
estimated via ab initio calculations of the
energetics, dielectric properties, and band  alignment
of bulk and thin-film oxides on Si (001).
With their large dielectric constants, stable 
and low-formation-energy interfaces, large valence offsets, and 
reasonable  (though not optimal) conduction offsets (electron
 injection barriers), zirconia and hafnia appear to have a
 considerable potential as gate oxides for Si electronics.
\end{abstract}
\maketitle
The performance needs of modern information technology are
 forcing Si-based ultra-large-scale-integrated (ULSI) 
devices into the domain of  nanometric dimensions. This
downscaling implies, among others, the effective continuing reduction
of the physical thickness of insulating gate oxide layers
in  CMOS (Complementary  Metal-Oxide-Semiconductor) devices.
Amorphous SiO$_2$, the natural oxide of Si technology, is now
nearing  its fundamental size  limits, with physical 
thicknesses currently down to 2 unit cells \cite{huff}.  This leads 
to uncomfortably large ($> 1$ A/cm$^2$) leakage currents 
and  increased failure probabilities. The main
reason for the strong  reduction of gate-oxide
thickness in  device downscaling 
is the need for increasing capacitances
in the  CMOS  conducting channel.
In a CMOS, the gate oxide layer dominates the series capacitance of
the channel.
 An increase in  capacitance  can be obtained reducing 
the dielectric thickness  $d/\varepsilon$ of the oxide layer, 
having physical thickness $d$ and relative dielectric constant
$\varepsilon$. Given its small dielectric constant, it is
understandable that SiO$_2$ as a gate oxide has emerged as one 
of the key bottlenecks in device donwscaling  \cite{huff,rev}. 

It thus appears
that,  if Moore's law \cite{moore} on ULSI circuit component density -
and hence circuit performance - is to remain valid in the next decade,
a replacement will have to be found for  silica as a gate insulator. 
The basic  selection criteria for such a
replacement are {\it i)} 
larger dielectric constant (``high-$\kappa$''), {\it ii)} interface 
band offsets to Si as 
large as or comparable to those of silica (especially the
electron injection barrier), {\it iii)} epitaxy on
 Si energetically not too costly, {\it iv)} thermodynamical stability 
in contact with Si. In this work we address the expected performance, in 
terms  of the above criteria, for the two important current candidates
\cite{huff,rev,huff2}
hafnia (HfO$_2$) and zirconia (ZrO$_2$) through
first-principles density-functional calculations of the
structure, energetics, thermodynamical stability, dielectric constants,
and band offsets of crystalline
hafnia and zirconia thin films epitaxially grown
on the (001) face of crystalline Si. We find
stable and moderate-cost interfaces, large dielectric constants,
and large band offsets, except for the electron injection barrier,
 estimated at  1 eV at most, appreciably lower than the Si/silica 
barrier.

Our density functional theory calculations in the
generalized gradient approximation \cite{pw91} use the VASP code
\cite{vasp} and  the ultrasoft  \cite{uspp} pseudopotentials
 provided therewith. Semicore states are treated as core for Hf and
Zr;  test calculations  done including the semicore as valence
using the  all-electron  PAW \cite{paw} method as implemented 
in VASP \cite{vasp}  confirmed the pseudopotential
results. Bulk optimizations were done   
in a 12-atom (conventional fcc or fct) cell,
while the interfaces are simulated by (001)-oriented 
oxide/Si superlattices contained in tetragonal cells of $c(2\times2)$ 
basal section, and in-plane lattice
constant $a_{\rm Si}$=5.461 \AA, our theoretical value for bulk
Si. Interface supercells contain around 50 atoms
depending on the local interface structure, with  9 layers (18 atoms)
for the  Si region, and 
typically 11 layers (e.g.  24 oxygen and 10 Zr atoms)
for the oxide region. The
 plane-wave basis cutoff is 350 eV; for the k-space summation 
we use 4$\times$4$\times$4 meshes for the
bulk and 4$\times$4$\times$1 meshes for the $z$-elongated 
interface supercells.
 
\paragraph*{Bulk and Si-epitaxial structure -- } 
 Bulk hafnia and zirconia 
were studied in the fluorite,  monoclinic, and
Si-epitaxial  structures. 
The lattice parameters for ZrO$_2$ are $a$=5.10 \AA\, for fluorite,
and  ($a$,$b$,$c$)=(5.186,5.255,5.351) \AA,
off-normal angle $\theta$=8.83$^{\circ}$  for monoclinic.  
For HfO$_2$, $a$=5.06 \AA\, for   fluorite, and
  ($a$,$b$,$c$)=(5.108,5.175,5.280) \AA, 
 off-normal angle $\theta$=8.80$^{\circ}$ for monoclinic.
The latter phase is favored over fluorite
 by 0.115 eV/formula unit for ZrO$_2$ and
by 0.248 eV/formula unit for HfO$_2$.
The results agree  with experiment and
with recent calculations \cite{vand-z,gonze,vand-h}.
The formation enthalpies $\Delta {\rm H}_{\rm ox}$ are
--11.52 eV and --10.74 for hafnia and zirconia respectively (close to 
experiment, as usual using GGA) compared to
--8.30 eV for silica: therefore  both oxides are stable
 in contact  with Si  with respect to the decomposition into silica
and metal. The same holds for the epitaxial phase discussed next, 
whose excess energy  is only about 0.2 eV/formula above the
monoclinic. 

\begin{figure}[h]
\epsfclipon
\epsfxsize=7cm
\epsffile{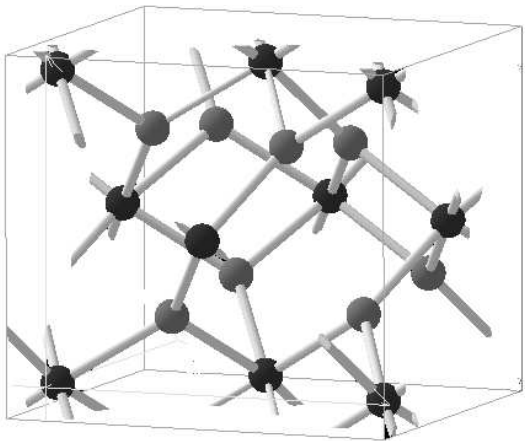}
\caption {Si-epitaxial structure of ZrO$_2$. Grey (black) atoms: O (Zr).}
\label{fig1}
\end{figure}

The tetragonal
Si-epitaxial  crystalline phase of each oxide was obtained imposing
the 
 in-plane lattice constant of Si, and adjusting the axial
 ratio and internal coordinates in the 12-atom conventional cells.
 The axial ratios $c$/$a_{\rm Si}$ are 
0.92 for ZrO$_2$ and 0.90 for HfO$_2$.
We verified by variable-cell damped dynamics \cite{vasp} that this
tetragonal bulk is stable against monoclinic distortions.
The
Si-epitaxial configuration, depicted in Fig. \ref{fig1} for ZrO$_2$,
 may be viewed as a
$z$-stacking of cation-anion  bilayers alternatingly oriented at 
90$^{\circ}$  
to each other, in which  {\it a)} metal cations are disposed in
dimerized (110)-like  rows (cation-cation distances within the rows
 3.4 and 4.2
 \AA\,  compared to 3.86 \AA\, ideally), and {\it b)}
oxygens  quadruplets, originally  square in fluorite, elongate to
 rhomboids along the (110) rows bending slightly sideways. 
The cation (anion) coordination decreases from 8 to 6 (from 4 to 3),
 in partial  analogy to the monoclinic structure \cite{vand-z}.

The elastic energy $E_{\rm elastic}^{\rm epi}$
 of the Si-epi distorted bulk  is
0.23 eV/formula or 5.87 meV/\AA$^3$ for ZrO$_2$, and 0.16
 eV/formula or 4.37 meV/\AA$^3$  for HfO$_2$
 with
respect to monoclinic bulk (i.e. both are slightly favored 
energetically over fluorite, whose occurrence is anyway barred by 
symmetry).  While substantial, these energies are comparable to those
 of order   $\sim$4 meV/\AA$^3$ involved  (for much smaller strains)
 in nitride semiconductor epitaxy  \cite{noi-sl}.   
As we now discuss, the knowledge of the
volume-specific epitaxial strain energy
enables us to extract an area-specific interface
energy, as well as to estimate the critical
pseudomorphic growth  thickness.

\paragraph*{Interface energetics and offsets --}

Assuming a $c$(2$\times$2) basal section, we investigated for both
materials several local structures and terminations
of oxide/Si (001) interfaces,  e.g.  Si/O,
Si/metal,  Si/metal-bilayer,  mixed Si-metal layer/O,
 mixed Si-metal layer/O with 50\%  vacancies. 
The starting configuration of the oxide portion of
the  interface superlattices is assembled 
using the optimized  Si-epi structure.
The supercell length and 
atomic positions are then reoptimized: the
 axial ratio remains unchanged, and 
 relaxations occur only in the first two
interface-neigboring layers.
The interface energy   can be expressed as the difference of
the energies $E_{\rm SL}$  
of the interface cell, and $E_{\rm bulk}$ of
 the corresponding bulk components, as
\begin{eqnarray}
E_{\rm form} &=& \frac{1}{2\, A} [E_{\rm SL} - E_{\rm bulk}] 
= \frac{1}{2\, A} [( 
2\, A\, \delta + n_{\rm Si} V_{\rm Si} E_{\rm Si} + \nonumber \\
&+ & n_{\rm ox} V_{\rm ox} E_{\rm ox})
-  (n_{\rm Si} V_{\rm Si} E_{\rm Si} + 
n_{\rm ox} V^{'}_{\rm ox} E^{'}_{\rm ox})] = \delta \nonumber
\end{eqnarray}
with $n$ the number of bulk units,  $V$, $V'$ and 
$E$, $E'$ the corresponding 
volumes  and energies per unit volume, $A$ the basal superlattice 
area. The formation energy per unit area,
 $\delta$, can be extracted unambiguously if
 the  oxide bulk energy is calculated in the same strain state 
as in the superlattice (Si remains unstrained), as 
in that case all volume-dependent terms drop off. Any
other choice of the bulk energies  inserts a volume
dependence  in the interface energy \cite{noi-sl}.  

The interface cell may be stoichiometric,
 metal- or oxygen-deficient depending on 
its local structure. Its formation  energy will therefore depend
on growth conditions, metal-rich ones favoring oxygen deficit, and
O-rich favoring oxygen excess. Theoretically,
 this is described by fixing the  chemical
potentials of the constituents. Here, only one
 potential -- e.g. oxygen's -- is independent:
 $\mu_{\rm O}=\mu_{\rm O_2}/2$ means  O-rich conditions,
and 
 $\mu_{\rm O}=\mu_{\rm O_2}/2 + \Delta {\rm H}_{\rm ox}/2$
 metal-rich ones.

\begin{table}
\caption{\label{tab1} 
Formation energies (eV/\AA$^2$) of,
and valence and conduction band offsets (eV) at
different Si (001)/oxide interfaces. The assumed growth
conditions  are indicated. The best offset/energetics
combinations are displayed in underlined  bold for metal-rich 
conditions, and bold for oxygen rich conditions.
All GW corrections are included.}
\begin{ruledtabular}
\begin{tabular}{c|c|ccr|ccr}
Material & $\rightarrow$  & &{HfO$_2$}  &   &   &{ZrO$_2$} &   \\
 \hline 
Interface $\downarrow$& Growth & VBO  &CBO  & E$_{\rm form}$  & VBO  &CBO & E$_{\rm form}$  \\
\hline
 Si/O      & O-rich & 4.14    & 0.47 &--0.16  & 4.08  & {\bf 0.72} &{\bf --0.21}   \\
Si-M/O     & stoich& 4.40  &0.19 & 0.17   & 4.18 & 0.62  & 0.12 \\
Si/M       & M-rich& 3.96  &0.65 &  0.12 &  4.72 & 0.08 & 0.07 \\
Si-M/O vac  & M-rich &   &  &  &  3.91  &  
\underline{\bf 0.89} & \underline{\bf --0.15}\\
Si/O vac  & stoich & 4.62 &--0.01 & 0.22 & 3.70 & 1.10 & 0.13\\
\end{tabular}
\end{ruledtabular}
\end{table}

\begin{figure}[h]
\epsfclipon
\epsfxsize=7cm
\epsffile{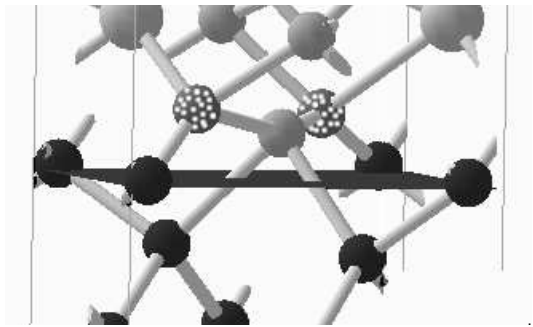}
\caption {The [mixed metal-Si]/[O 50\% vacant]
interface (black: Si; grey: metal; white-dotted: O) .}
\label{fig2}
\end{figure}

The formation  energies of the various interfaces
are listed in Table \ref{tab1}. The standard Si-O
interface is favored  in O-rich growth conditions.
 In metal-rich conditions, the preferred structure 
is the mixed Si-metal to 50 \% vacant oxygen layer interface
depicted in Fig. \ref{fig2}, which remarkably is the same
as was recently obtained \cite{surfdyn} in all-electron ab-initio 
molecular dynamics  simulations of metal deposition on, and 
oxidation of, Si (001).  Notably, the two favored interfaces have 
large {\it negative} formation energies (referred, we remind, to the
pre-strained bulk). This  energetic gain in interface formation
will be counterbalanced by the excess energy of the film's upper
surface, and by the build-up of epitaxial elastic  energy in 
the growing layer. An estimate of the critical thickness $t_c$ for
pseudomorphic growth over an area $A$ then results from
$$
A E_{\rm form} + A t_c\, E_{\rm elastic}^{\rm epi} + A E_{\rm surf} = 0,
$$
using which we predict that  crystalline zirconia and hafnia thin films
 should grow pseudomorphically on Si (001): indeed, using
 our calculated values for, e.g., zirconia, and the 
GGA surface-energy estimate for the  tetragonal phase
  $E_{\rm surf}\simeq$ 0.05 eV/\AA$^2$ \cite{carter},
we obtain   $t_c\sim$18 \AA\, and 27 \AA\,  for metal- and
oxygen-rich conditions respectively.
 The poly-Si gate, forming a Si/oxide interface
in the place of a free oxide surface, should further stabilize 
the structure.

The interface  band  offsets are evaluated for each interface
using the standard 'bulk-plus-lineup' procedure 
\cite{bald}, expressing the valence offset (VBO)
as the sum of the interface potential lineup and
the valence-band-top differences of the separately-considered bulks.
  The conduction band offsets, hence 
the electron injection barriers,  is estimated as 
CBO = E$_{\rm gap}^{\rm oxide}$ --  E$_{\rm gap}^{\rm Si}$ -- VBO.  
The gap of Si  is taken to be 1.1 eV; for both oxides, we use our 
GGA gaps corrected with the GW data of Ref. \cite{kralik} for 
ZrO$_2$,  namely 5.9 eV and 5.7 eV for zirconia and
hafnia respectively. These values are close to experiment for
hafnia, and near the bottom of the (large) experimental range 
for zirconia. We neglect spin-orbit
corrections, which should be well below
$\pm$0.1 eV as the valence states are oxygen-derived.  
We do include, instead, the quasiparticle corrections to the bulk
 valence-band edges  at the GW level: this is essential since these
 corrections are of order $\sim$1 eV  in oxides compared to typical
 $\sim$0.1 eV in semiconductors.
 We apply  to the VBOs an overall correction
 of --1.08 eV, resulting from the  --0.15 eV correction 
\cite{zhu13} for Si   and the  --1.23 eV correction \cite{kralik} for 
ZrO$_2$. Using the latter for both oxides introduces some uncertainty in
the HfO$_2$ results, but unfortunately no GW data are currently 
available for hafnia.

In Table \ref{tab1} we report 
the  predicted VBOs and CBOs. Qualitatively,
 VBOs cluster around 4 eV, with appreciable  structure dependence, 
and  CBOs are in the range 0 to 1 eV. Interestingly, for zirconia
 the   energy-wise most favorable  structures have some of
 the largest conduction offsets. The high-end CBOs, $\sim$ 1 
eV, are smaller than, but comparable to, the 1.4-1.5 eV 
estimates by Robertson \cite{robertson}, who
 used a simple charge-neutrality-level model at the empirical 
tight-binding model.

\paragraph*{Dielectric constants --} The lattice contribution to
the dielectric tensor has been calculated
 for both oxides in the fluorite, monoclinic, and Si-epitaxial
structures. 
 We  used a standard formalism to evaluate the zero-frequency
dielectric constant \cite{vand-z} via the
frequencies of zone-center IR-active modes and the transverse 
dynamical charges. The vibrational modes are calculated diagonalizing the  
zone-center dynamical matrix 
\mbox{--$\partial F^{\alpha}_{i}/\partial u^{\beta}_{j}$}, obtained 
differentiating by  
centered finite-differences   (with displacements of 0.1 \AA) 
 the Hellmann-Feynman force component $\alpha$  on atom $i$
with respect to the displacement of atom $j$ along direction $\beta$.
The dynamical charges are likewise obtained by 
finite-difference differentiation of the Berry-phase \cite{berry}
 polarization with respect to atomic displacements (of typically
 0.05 \AA).

\begin{table}[h]
\caption{\label{tab2} Lattice 
dielectric tensor for
  fluorite, monoclinic, and Si-epitaxial  XO$_2$
(the small off-diagonal 
elements for the monoclinic are not displayed for clarity),
calculated using 
the dynamical charge tensor of fluorite.
  $\varepsilon^{\rm ave}_{\rm lat}$  is
the  orientational average
 measured by series capacitance
in polycristalline layers, and obtained as
 $3/\varepsilon^{\rm ave} =  
 1/\varepsilon^{xx} +
 1/\varepsilon^{yy} +
 1/\varepsilon^{zz}$,
}
\begin{ruledtabular}
\begin{tabular}{lccc|c}
 &  $\varepsilon^{xx}_{\rm lat}$ 
 &  $\varepsilon^{yy}_{\rm lat}$ 
 &  $\varepsilon^{zz}_{\rm lat}$ 
 &  $\varepsilon^{\rm ave}_{\rm lat}$ \\
\hline
HfO$_2$ fluorite & 27.8 & 27.8  & 27.8 &  27.8 \\
HfO$_2$ monoclinic & 17.5 &     15.7&            12.4& 14.9 \\
HfO$_2$ Si-epi   & 27.6 & 18.6 & 24.5 & 22.9 \\
\hline
ZrO$_2$ fluorite &  30.5& 30.5 & 30.5 & 30.5 \\
ZrO$_2$ monoclinic & 24.7 &     18.3 &        14.6 & 18.4 \\
ZrO$_2$ Si-epi & 22.5 & 71.5 & 44.9 & 37.0 \\
\end{tabular}
\end{ruledtabular}
\end{table}

Since the epi-oxides were optimized without  
constraints, they have no symmetry of practical use.
The calculation of the full
dynamical charge tensor for all atoms in the complex epitaxial 
(as well as the monoclinic) structure is thus rather demanding,
and currently in progress. 
In  Table \ref{tab2} we give   estimates of the
diagonal elements of the lattice dielectric tensor obtained  using the 
dynamical charge tensor of the fluorite phase, which is
diagonal and isotropic,  and
calculated to be  $Z^*_{\rm Hf}$=5.20 and
$Z^*_{\rm O}$=--2.60 for HfO$_2$,
 $Z^*_{\rm Zr}$=5.50 and  $Z^*_{\rm O}$=--2.75
for ZrO$_2$. Of course, smaller  dynamical charges 
such as found in monoclinic phases \cite{vand-z,vand-h} will
 decrease the dielectric constant,  especially the
$zz$ component. Using the monoclinic cation charge tensors
of Refs. \cite{vand-z,vand-h} and imposing the Friedel sum rule
to obtain an average anion charge tensor,
we estimated $\varepsilon_{zz}$ to be 9.9 and 11.9 in  
monoclinic 
HfO$_2$ and  ZrO$_2$ respectively, in fair agreement with
previous results.
Along with our fluorite values, also in good agreement with previous
calculations, this gives us confidence on the reliability of our 
procedure.

\begin{figure}[h]
\epsfclipon
\epsfxsize=7cm
\epsffile{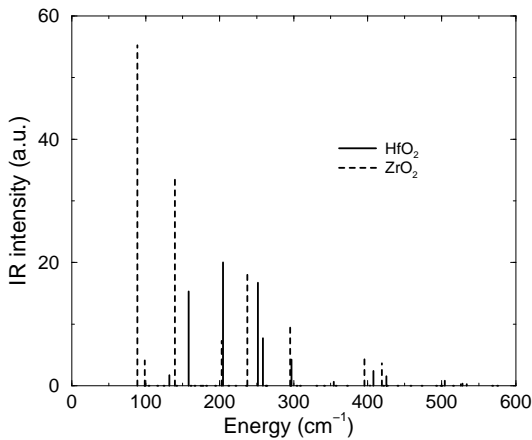}
\caption {Orientationally
 averaged IR intensity spectrum (mode dielectric constants)
of Si-epitaxial HfO$_2$ (solid) and ZrO$_2$ (dashed).}
\label{epsm-z}
\end{figure}

With reference to Table \ref{tab2},
for hafnia we find   a  reduction in dielectric constant
compared to fluorite both in the  Si-epi and monoclinic phases,
though the latter is rather more  dramatic, with a more than twofold
decrease, in agreement with previous calculations \cite{vand-h}.
For zirconia, we also find a similar, approximately twofold 
reduction of  the monoclinic dielectric tensor compared to fluorite; 
notably, though,  a drastic enhancement is found
 in the Si-epitaxial phase. This results  from the large 
IR intensity of modes  at about 90 cm$^{-1}$ to 140 cm$^{-1}$, as
can be seen in  Fig.\ref{epsm-z},
which reports the mode dielectric constants \cite{vand-z} for
both materials in the Si-epi phase.  The  two lower-energy modes 
for zirconia (dashed lines)  contribute mostly to the $yy$ component, 
the third to the $zz$ component. The pronounced softness of 
Si-epi zirconia is 
presumably due to the backfolding of zone-border ($X$-point) modes. 

We carefully checked against 
artifacts by accurately reoptimizing  structures and
 repeating phonon calculations for different displacements. We
 are confident in our procedure also in view of the results for
the other phases. The single zone-center IR-active mode of fluorite is
  $\omega$=230 cm$^{-1}$  for HfO$_2$ and $\omega$=258
cm$^{-1}$ for ZrO$_2$;  for the latter this agrees with recent  
predictions \cite{vand-z,gonze}, for the former  the
frequency is 20\% lower than in Ref. \cite{vand-h}. We
checked that the same results are obtained (within 0.5\% for 
the lattice constant and  2\% for the frequency) with the 
all-electron PAW method with  valence semicore \cite{vasp,paw}.
The details of the vibrational spectrum of the epi and monoclinic phases
will be reported elsewhere, but we note in passing that the results for
the  monoclinic are close to previous reports \cite{vand-h}.

In conclusion, the picture of
 zirconia and hafnia as Si-gate oxides as it emerges from this
work is rather encouraging, certainly
so from the dielectric and epitaxy-energetic standpoints.
The results on the electron injection barriers are
partly disappointing, as the 
electron injection barrier is much smaller that at silica/Si 
interfaces. While insufficient for hot 
electrons, the barrier should be still acceptable  for standard
two-dimensional inversion layers, whose energy
levels  are at about 100 meV above the interface triangular-well 
bottom \cite{andofowlerstern}.

We thank P. Bl\"ochl and C. F\"orst for communicating their surface
results  for Zr-O coadsorption on Si (001) prior to publication.
We are grateful to A. Pasquarello and P. Pavone for helpful suggestions.
This work was supported in part by the European Union (INVEST project),
and by the Parallel Supercomputing Initiative of
 Istituto Nazionale per la Fisica della Materia.

\end{document}